\documentclass[aps,pra,showpacs,nofootinbib,twocolumn]{revtex4}
\usepackage{latexsym}
\usepackage{amsfonts}
\usepackage{amsbsy}
\usepackage{mathrsfs}

\begin{document}
\title{Symplectic quantization, inequivalent quantum theories, and
Heisenberg's principle of uncertainty}
\date{\today}

\author{Merced Montesinos\footnote{Associate Member of the Abdus Salam
International Centre for Theoretical Physics, Trieste, Italy.}}
\email{merced@fis.cinvestav.mx} \affiliation{Departamento de F\'{\i}sica,
Centro de Investigaci\'on y de Estudios Avanzados del I.P.N., Av. I.P.N. No.
2508, 07000 Ciudad de M\'exico, M\'exico}

\author{G. F. Torres del Castillo}
\email{gtorres@fcfm.buap.mx} \affiliation{Departamento de F\'{\i}sica
Matem\'atica, Instituto de Ciencias,\\
Universidad Aut\'onoma de Puebla, 72570 Puebla, Pue., M\'exico}
\begin{abstract}
We analyze the quantum dynamics of the non-relativistic two-dimensional
isotropic harmonic oscillator in Heisenberg's picture. Such a system is taken
as toy model to analyze some of the various quantum theories that can be built
from the application of Dirac's quantization rule to the various symplectic
structures recently reported for this classical system. It is pointed out that
that these quantum theories are inequivalent in the sense that the mean values
for the operators (observables) associated with the same physical classical
observable do not agree with each other. The inequivalence does not arise from
ambiguities in the ordering of operators but from the fact of having several
symplectic structures defined with respect to the same set of coordinates. It
is also shown that the uncertainty relations between the fundamental
observables depend on the particular quantum theory chosen. It is important to
emphasize that these (somehow paradoxical) results emerge from the combination
of two paradigms: Dirac's quantization rule and the usual Copenhagen
interpretation of quantum mechanics.
\end{abstract}
\pacs{03.65.Ta, 03.65.Ca} \maketitle
\section{Introduction}
The usual textbook treatment of the Hamiltonian formulation of dynamical
systems consists in writing the equations of motion
\begin{eqnarray}\label{zero}
{\dot q}^i = f^i (q,p) \, , \quad {\dot p}_i = g_i (q,p) \, ,
\end{eqnarray}
for autonomous systems in the form \cite{Berndt,Olver}
\begin{eqnarray}
\label{HamEq} {\dot q}^i & = & \frac{\partial H}{\partial p_i} \, , \quad
{\dot p}_i = -\frac{\partial H}{\partial q^i} \, , \quad i,j=1,2,..., n \, ,
\end{eqnarray}
where $H$ is ``the Hamiltonian of the system," the variables $(q^i,p_i)$ are
canonically conjugate to each other in the sense that
\begin{eqnarray}\label{usualPB}
\{ q^i , q^j \} =0\, , \quad \{ q^i , p_j \}= \delta^i_j\, , \quad \{ p_i ,
p_j \}=0 \, ,
\end{eqnarray}
with $\{ , \}$ the Poisson bracket defined by
\begin{eqnarray}\label{usualPBII}
\{ f , g \}  = \frac{\partial f}{\partial q^i} \frac{\partial g}{\partial p_i}
- \frac{\partial f}{\partial p_i} \frac{\partial g}{\partial q^i} \, ,
\end{eqnarray}
where from now on Einstein's sum convention over the contracted indices is
understood.

If the classical system admits the Hamiltonian formulation previously
mentioned, then the {\it standard recipe} to go from its classical to its
quantum dynamics from the canonical point of view consists in finding an
irreducible representation for the fundamental operators which satisfy the
Heisenberg-Born-Jordan relations or simply canonical commutation relations
\cite{Dirac,Heis,Heis2}
\begin{eqnarray}\label{KeyEq}
\left [ {\widehat q}^i , {\widehat q}^j \right ] =0 \, ,\quad \left [
{\widehat q}^i , {\widehat p}_j \right ] =i \hbar \delta^i_j \, , \quad \left
[ {\widehat p}_i , {\widehat p}_j \right ]=0 \, ,
\end{eqnarray}
which are the quantum version of Eqs. (\ref{usualPB}). In Eq. (\ref{KeyEq}),
the ``hat" over each symbol indicates the operator corresponding to the
variable under consideration and the square bracket $\left [ , \right ]$
indicates the commutator of operators. For most of dynamical systems with a
finite number of degrees of freedom the specific representation of these
operators does not matter, on account of the Stone-von Neumann theorem
\cite{Stone} [nevertheless, an exception where the theorem does not apply is
the system of a ``particle in a box"]. It is important to emphasize that the
standard procedure to go from the classical to the quantum realm, known as
canonical quantization \cite{Dirac}, is not completely free of ambiguities.
Among them one has the choice of the measures on the several Hilbert spaces
involved and sometimes some ambiguities in the ordering of the product of
operators. Even though these ambiguities are important, they are not relevant
for the present discussion and they are mentioned just for completeness in the
description of canonical quantization.

Coming back to the classical dynamics and before mentioning the ideas
developed in this paper, it is convenient to remind the reader that the
equations of motion (\ref{HamEq}) can be put in the form
\begin{eqnarray}\label{cinco}
{\dot x}^{\mu} = \omega^{\mu\nu} \frac{\partial H}{\partial x^{\nu}}\, , \quad
\mu,\nu=1,2,...,2n\, ,
\end{eqnarray}
with $(x^{\mu})=(q^1,q^2,...,q^n,p_1,p_2,...,p_n)$ and
\begin{eqnarray}\label{seis}
\left ( \omega^{\mu\nu} \right ) = \left (
\begin{array}{cc}
0 & I \\
-I & 0
\end{array}
\right )\, ,
\end{eqnarray}
where $0$ and $I$ are $n\times n$ matrices \cite{Berndt}. Also, Eq.
(\ref{usualPBII}) acquires the form
\begin{eqnarray}\label{newPB}
\{ f , g \} = \frac{\partial f}{\partial x^{\mu}} \omega^{\mu\nu}
\frac{\partial g}{\partial x^{\nu}} \, ,
\end{eqnarray}
from which Eq. (\ref{usualPB}) can be rewritten as
\begin{eqnarray}
\{ x^{\mu} , x^{\nu} \} = \omega^{\mu\nu} \, .
\end{eqnarray}
Thus, from this perspective, the coordinates $(x^{\mu})$ locally label the
points of the phase space $\Gamma$ of the system where dynamics takes place,
the Hamiltonian $H$ is a real function defined on $\Gamma$, and the definition
of the Poisson bracket is equivalent to the introduction of a symplectic
structure $\omega= \frac12 \omega_{\mu\nu} d x^{\mu} \wedge d x^{\nu}$ on the
phase space, where the matrix $(\omega_{\mu\nu})$ is the inverse matrix of
$(\omega^{\mu\nu})$. The 2-form $\omega$ is non-degenerate, i.e.,
$\omega_{\mu\nu} v^{\nu}=0$ implies $v^{\mu}=0$ which means that there exists
the inverse matrix $(\omega^{\mu\nu})$. Also, $\omega$ is closed, i.e.,
$\partial_{\mu} \omega_{\nu\gamma} + \partial_{\nu} \omega_{\gamma\mu} +
\partial_{\gamma} \omega_{\mu\nu}=0$ which is equivalent to the fact that the
Poisson bracket satisfies the Jacobi identity \cite{Berndt,Olver}.

Therefore, it is clear that the symplectic geometry is the geometric structure
underlying the Hamiltonian formulation of mechanics \cite{Berndt,Olver}.
Moreover, Eqs. (\ref{cinco}) are covariant in the sense that they maintain
their form if the canonical coordinates are replaced by a completely arbitrary
set of coordinates in terms of which $(\omega^{\mu\nu})$ need not be given by
Eq. (\ref{seis}). It should be remarked that even in the standard formulation
of Lagrangian or Hamiltonian mechanics one always has the possibility of using
completely arbitrary coordinates in the configuration or in the phase space;
the usual procedure consists in finding first the expression for the
Lagrangian or the Hamiltonian function making use of an inertial reference
frame and then make the desired coordinate transformation.

In a similar way, one can retain the original coordinates $(q^i,p_i)$ and
still write the original equations of motion (\ref{zero}) in the Hamiltonian
form (\ref{cinco}), but now employing alternative symplectic structures
$\omega^{\mu\nu}(x)$, distinct to that given in Eq. (\ref{seis}), and by
taking as Hamiltonian any real function on $\Gamma$ which is a constant of
motion for the system. This means that the writing of the equations of motion
of a dynamical system in Hamiltonian form is {\it not} unique [see Sect.
\ref{Intro}]. It is pretty obvious that any description of the dynamics for a
given classical system from the symplectic point of view is mathematically and
physically acceptable.

However, it is {\it a priori} far from being obvious whether or not the
various quantum theories emerging from the combination of Dirac's quantization
rule
\begin{eqnarray}\label{Drule}
[ {\widehat f} , {\widehat g} ] & = & i \hbar \widehat{ \{ f, g \}} \, ,
\end{eqnarray}
with alternative symplectic structures are mathematically and physically
equivalent to each other in the generic case. Again, in Eq. (\ref{Drule}), the
``hat'' over the each symbol indicates the operator corresponding to the
classical variable under consideration. Therefore, $\widehat{ \{ f, g \}}$ is
the operator corresponding to $\{ f , g \}$. In particular, the combination of
these two ingredients gives rise to the following questions: what are the
consequences in the quantum theory of choosing alternative symplectic
structures on the phase space of the theory when the pairs $(q^i,p_i)$ are not
necessarily canonical ones from the very beginning? Is it possible in such
cases to build a mathematically ``consistent" quantum theory? If the answer is
in the affirmative, does it make sense physically? In this paper, we are going
to try to answer these kinds of questions.

At first sight it might appear that this way of approaching quantum mechanics
is the one of geometric quantization \cite{Woodhouse}. Nevertheless, there,
people frequently choose a symplectic structure in such a way that $(q^i,p_i)$
are canonical coordinates to start with the quantization programme.

Here, as we mentioned, we are not interested in keeping $(q^i,p_i)$ as
canonical coordinates but exactly the other way around, we want to analyze the
quantum theories that emerge from Dirac's quantization rule (\ref{Drule}) when
alternative symplectic structures (defined with respect to the same set of
coordinates of phase space) are taken into account. To investigate this point,
the quantum dynamics of the two-dimensional isotropic harmonic oscillator is
analyzed in this paper. In particular, it is shown that several quantum
theories can consistently be built from alternative symplectic structures
associated with the {\it same} classical system and that the corresponding
quantum theories are not equivalent in the sense that the expectation values
for the operators (observables) associated to the same physical entity do not
agree with each other in all these quantum theories. Moreover, it is shown
that Heisenberg's uncertainty principle in the usual way that it is normally
stated does not hold. In our opinion, these results are just a reflection of
the fact that the notions involved in quantum mechanics are not expressed in a
``covariant way" but they are tied to the case when $(q^i,p_i)$ are canonical
coordinates as we discuss in Sects. \ref{quantum}, \ref{Hup}, and
\ref{Remarks}.

\section{Freedom in the symplectic description of classical dynamics}\label{Intro}
Before going into the quantum theory, it is convenient to review the classical
dynamics of the non-relativistic two-dimensional isotropic harmonic oscillator
which will be used as toy model to study the consequences on the quantum
theory of choosing symplectic structures alternative to the usual one. The
dynamics of this system is given by the equations of motion
\begin{eqnarray}
{\dot x} & = & \frac{p_x}{m}\, , \quad \quad\quad \quad {\dot y}
=\frac{p_y}{m}\, , \nonumber\\
{\dot p}_x & = & - m \omega^2 x\, , \quad \,\,\, {\dot p}_y = - m\omega^2 y \,
, \label{THO}
\end{eqnarray}
where the dot ``$\cdot$" stands for the time derivative with respect to the
Newtonian time $t$, $m$ is the mass and $\omega$ the angular frequency. The
solution to the equations of motion (\ref{THO}) is
\begin{eqnarray}\label{Sol}
x & = & x_0 \cos{\omega t} + \frac{p_{x0}}{m\omega}\sin{\omega t}
\, , \nonumber\\
p_x & = &  - m \omega x_0 \sin{\omega t} + p_{x0} \cos{\omega t}
\, ,\nonumber\\
y & = & y_0 \cos{\omega t} + \frac{p_{y0}}{m\omega}\sin{\omega t}
\, , \nonumber\\
p_y & = &  - m \omega y_0 \sin{\omega t} + p_{y0} \cos{\omega t} \,
,
\end{eqnarray}
where $x_0 = x(t=0)$, $y_0 = y(t=0)$, $p_{x0}= p_x (t=0)$, and $p_{y0}= p_y
(t=0)$ are the initial data (at $t=0$) of the dynamical variables for the
system.

{\it Usual viewpoint of symplectic dynamics}. The standard treatment found in
textbooks about the writing of the equations of motion (\ref{THO}) in the
Hamiltonian formalism (\ref{cinco}) is:

0) to consider that the points of the phase space $\Gamma=\mathbb{R}^4$ are
labelled by the coordinates $(x^{\mu})=(x^1,x^2,x^3,x^4)=(x,y,p_x,p_y)$ in
such a way that $(x,p_x)$ and $(y,p_y)$ are canonical pairs. From this point
of view, equations (\ref{THO}) can be put in the form (\ref{cinco}) with
\begin{eqnarray}\label{CanSE}
(\omega^{\mu\nu}) = \left ( \begin{array}{cccc}
0 & 0 & 1 & 0 \\
0 & 0 & 0 & 1 \\
-1 & 0 & 0 & 0 \\
0 & -1 & 0 & 0
\end{array} \right ) \, ,
\end{eqnarray}
and
\begin{eqnarray}\label{energy}
\quad H= S_0 := \frac12 \left ( \frac{(p_x)^2}{m} + m\omega^2 x^2 +
\frac{(p_y)^2}{m} + m\omega^2 y^2 \right ) \, ,
\end{eqnarray}
or, equivalently, the non-vanishing Poisson brackets
are
\begin{eqnarray}\label{CanPB}
\{ x , p_x \}_0 = 1\, , \quad \{ y , p_y \}_0 = 1 \, ,
\end{eqnarray}
which is the same as
\begin{eqnarray}\label{AAA}
\omega_0= dp_x \wedge dx + d p_y \wedge d y \, .
\end{eqnarray}

{\it Alternative viewpoints of symplectic dynamics}. As explained in Refs.
\cite{Ger0,Ger1}, it is not mandatory to interpret $(x,p_x)$ and $(y,p_y)$ as
if they were {\it per se} canonical coordinates, and many other choices of the
pair $(\omega, H)$ where $\omega$ is a symplectic structure and $H$ is a
Hamiltonian are allowed. The following four pairs were introduced in Ref.
\cite{Ger1}:

i) the equations of motion (\ref{THO}) can be put in a Hamiltonian form
(\ref{cinco}) by taking $(x^{\mu})=(x^1,x^2,x^3,x^4)=(x,y,p_x,p_y)$,
\begin{eqnarray}
(\omega^{\mu\nu}) = \left ( \begin{array}{cccc}
0 & 0 & 0 & 1 \\
0 & 0 & 1 & 0 \\
0 & -1 & 0 & 0 \\
-1 & 0 & 0 & 0
\end{array} \right ) \, ,
\end{eqnarray}
and
\begin{eqnarray}
\quad H= S_1 := \frac{p_x p_y}{m} + m \omega^2 x y \, ,
\end{eqnarray}
or, equivalently, the non-vanishing Poisson brackets are
\begin{eqnarray}\label{CanPBi}
\{ x , p_y \}_1 = 1\, , \quad \{ y , p_x \}_1 =1 \, ,
\end{eqnarray}
which is the same as
\begin{eqnarray}
\omega_1 = d p_y \wedge d x + d p_x \wedge dy
\end{eqnarray}
[cf. Eqs. (\ref{CanPB}) and (\ref{AAA})].

ii) the equations (\ref{THO}) can also be obtained from
$(x^{\mu})=(x^1,x^2,x^3,x^4)=(x,y,p_x,p_y)$,
\begin{eqnarray}
(\omega^{\mu\nu}) = \left ( \begin{array}{cccc}
0 & 0 & -1 & 0 \\
0 & 0 & 0 & 1 \\
1 & 0 & 0 & 0 \\
0 & -1 & 0 & 0
\end{array} \right ) \, ,
\end{eqnarray}
and
\begin{eqnarray}
\quad H= S_2 := \frac{(p_y)^2 -(p_x)^2}{2m} + \frac12 m \omega^2 (y^2 - x^2)\,
,
\end{eqnarray}
or, equivalently, the non-vanishing Poisson brackets
are
\begin{eqnarray}\label{CanPBii}
\{ x , p_x \}_2 = -1 \, , \quad \{ y , p_y \}_2 =1 \, ,
\end{eqnarray}
which is the same as
\begin{eqnarray}
\omega_2= - d p_x \wedge dx + d p_y \wedge d y
\end{eqnarray}
[cf. Eqs. (\ref{CanPB}) and (\ref{AAA})].

iii) similarly, the equations of motion (\ref{THO}) can be gotten from
$(x^{\mu})=(x^1,x^2,x^3,x^4)=(x,y,p_x,p_y)$,
\begin{eqnarray}
(\omega^{\mu\nu}) = \left ( \begin{array}{cccc}
0 & -\frac{1}{m\omega} & 0 & 0 \\
\frac{1}{m\omega} & 0 & 0 & 0 \\
0 & 0 & 0 & -m\omega \\
0 & 0 & m\omega  & 0
\end{array} \right ) \, ,
\end{eqnarray}
and

\begin{eqnarray}
\quad H= S_3 := \omega (x p_y - y p_x )\, ,
\end{eqnarray}
or, equivalently, the non-vanishing Poisson brackets
are
\begin{eqnarray}\label{CanPBiii}
\{ x , y \}_3 = - \frac{1}{m\omega} \, , \quad \{ p_x , p_y \}_3 = -m\omega \,
,
\end{eqnarray}
which is the same as
\begin{eqnarray}
\omega_3 = m \omega d x \wedge d y + \frac{1}{m\omega} d p_x \wedge d
p_y
\end{eqnarray}
[cf. Eqs. (\ref{CanPB}) and (\ref{AAA})].

Some remarks follow: 0) it is important to recall that the alternative
Hamiltonians are just constants of motion, that do not correspond nor they are
required to correspond to the energy of the system, as it can be easily
verified combining each Hamiltonian function with the Poisson bracket proposed
in each case or by using the equations of motion (\ref{THO}) directly.
Moreover, the energy is conserved in each case because it is a constant of
motion\footnote{A real function $f$ defined on the phase space $\Gamma$ is a
constant of motion if and only if $df/dt=0$. Therefore, to check if a function
$f$ is a constant of motion one needs to use the equation of motion only
without having to choose a particular Hamiltonian $H$ and its corresponding
symplectic structure $\omega$. Of course, if one makes a choice of the pair
$(H, \omega)$, then one can also use this knowledge to check it.}, 1) the
equations of motion do {\it not} uniquely determine a single pair $(\omega,
H)$ formed by a symplectic structure $\omega$ and a Hamiltonian $H$. In the
present case, the triples $(\Gamma=\mathbb{R}^4, \omega=\omega_{\mu}, H=
S_{\mu})$, $\mu=0,1,2,3$ give rise by means of Eq. (\ref{cinco}) to the same
equations of motion (\ref{THO}). Therefore, phrases involving an absolute
connotation like ``{\it the} Hamiltonian of the dynamical system" are not
correct because there is {\it not} a single Hamiltonian for a dynamical
system, rather, there are many of them \cite{Ger3,MonPRA03}\footnote{Note,
that the alternative symplectic matrices $(\omega^{\mu\nu})$ of the cases i),
ii), and iii) are {\it not} obtained from the matrix $(\omega^{\mu\nu})$ of
the case 0) by making the matrix product of the later by a matrix $K$, i.e.,
to write the equations of motion in a Hamiltonian form it is not required that
the alternative symplectic matrices $(\omega^{\mu\nu})$ are obtained by making
the matrix product of (\ref{seis}) by another matrix $K$.}. In addition, to
state that $x$ and $p_x$ (same for $y$ and $p_y$) ``do not commute
classically'' is, on account of the previously displayed symplectic
structures, also not correct because the commutation or not (in the Poisson
bracket sense) is not something intrinsic to the variables $x$ and $p_x$ but
it depends on the symplectic structure chosen\footnote{Moreover, to state that
``dynamical variables referring to different degrees of freedom do always
commute" is not correct. The consequences of this in the quantum theory and
its relationship with Heisenberg's principle of uncertainty will be discussed
later in this paper.} , 2) the fact of having several symplectic structures
$\omega_{\mu}$ should {\it not} be interpreted as a reflection of Darboux's
theorem \cite{Olver}, which applies once a symplectic 2-form has been defined
on a manifold of even dimension. Here, there is no such a {\it fixed}
symplectic structure from the very beginning, rather, one is defining four
{\it alternative} symplectic structures from the very beginning, 3) it must be
emphasized that even if the symplectic structure $\omega$ were fixed to be
(\ref{CanSE}), there would still be an ambiguity in the definition of the
Hamiltonian $H$, a constant $a$ might be added to $H$ to get a new Hamiltonian
$H+a$ . The converse is also true: if the Hamiltonian $H$ were fixed to be
(\ref{energy}), there would still be several ways of choosing the symplectic
structure $\omega$ in such a way that these choices, via Eq. (\ref{cinco}),
reproduce the equations of motion (\ref{THO}) [see Refs. \cite{Ger0,Ger1} for
more details], 4) note also that the difference among the several symplectic
structures is {\it not} a change of coordinates, the coordinates
$(x^{\mu})=(x^1,x^2,x^3,x^4)=(x,y,p_x,p_y)$ that label the points of the phase
space $\Gamma=\mathbb{R}^4$ are the {\it same} in {\it all} cases, what
changes is the choice of the pair $(\omega,H)$ formed by a symplectic
structure $\omega$ and a Hamiltonian $H$, 5) note that the Hamiltonian $S_0$
is bounded from below while the Hamiltonians $S_i$, $i=1,2,3$ are {\it not},
6) the symplectic structure of the case iii) implies classically a
non-commutativity between the coordinates $(x,y)$ and between the momenta
$(p_x,p_y)$.

By using (\ref{Sol}) it is possible to compute the corresponding symplectic
structures on the {\it physical} phase space $\Gamma_{phys}$ whose points are
labelled by the coordinates $(x_0,y_0,p_{x0},p_{y0})$. One gets
\begin{eqnarray}
\Omega_0 & = & d p_{x0} \wedge d x_0 + d p_{y0} \wedge d y_0\, , \nonumber\\
\Omega_1 & = & d p_{y0} \wedge d x_0 +  d p_{x0} \wedge d y_0 \, , \nonumber\\
\Omega_2 & = & - d p_{x0} \wedge d x_0 + d p_{y0} \wedge d y_0 \, ,
\nonumber\\
\Omega_3 & = & m\omega d x_0 \wedge d y_0 + \frac{1}{m\omega} d p_{x0} \wedge
d p_{y0} \, ,
\end{eqnarray}
respectively. Obviously
\begin{eqnarray}\label{family}
\Omega_{\mu} & = & (\phi_t)^{\ast} \omega_{\mu} \, , \quad
\mu=0,1,2,3,
\end{eqnarray}
with $\phi_t : \Gamma_{phys} \rightarrow \Gamma$ given by Eq. (\ref{Sol}),
i.e.,
\begin{eqnarray}\label{Final}
(\Gamma_{phys}, \Omega_{\mu}) \stackrel{\phi_t}{\longrightarrow}
(\Gamma,
\omega_{\mu}) \, , \quad \mu=0,1,2,3 \, .
\end{eqnarray}
Thus, even when the evolution in $t$ is a ``canonical transformation" there
exists to our disposal the freedom to choose the symplectic structure in the
target (in $\Gamma$) with the corresponding symplectic structure on the source
(in $\Gamma_{phys}$) with respect to which the ``abstract transformation"
given in (\ref{Sol}) becomes canonical [see Eq. (\ref{Final})].

In summary, there are many ways of making the description of classical
dynamics from the symplectic viewpoint, we have just listed four of them, and
all of these choices are mathematically and physically allowed. The reader
interested in the description of the non-relativistic two-dimensional harmonic
oscillator (as well as of any other dynamical system with first class
constraints only) from the parameterized point of view (which is also
covariant in the sense that the Newtonian time $t$ is treated on the same
footing as the other configuration variables) can see Ref. \cite{MonMon}, in
particular if he/she wants to understand the consequences on the constraints
formalism of the fact of having various symplectic structures (with respect to
the same set of coordinates) on the extended, on the constraints surface, and
on the reduced phase spaces associated with the same dynamical system.

\section{Inequivalent quantum theories}\label{quantum}
In the previous section, several forms of describing the classical dynamics of
the system (\ref{THO}) from a symplectic point of view were displayed. Now the
idea is to explore, in the framework of {\it symplectic
quantization}\footnote{We think that it is more appropriate to use the term
{\it symplectic quantization} instead of {\it canonical quantization} when, as
in the present cases, $(x,p_x)$ and $(y,p_y)$ are not always canonical
pairs.}, the quantum theories that emerge from each of these symplectic
structures under consideration. It will be shown that, in the context of the
so-called Copenhagen interpretation, the quantum theories are not equivalent
in the sense that the mean values of the operators (observables) associated to
the same physical classical entity do not agree.

The description of the quantum dynamics for the system will be given in the
Heisenberg picture. Thus, the quantum mechanical relations analogous to those
given in Eq. (\ref{Sol}) are given by
\begin{eqnarray}\label{QSol}
{\widehat x} (t) & = & {\widehat x}_0 \cos{\omega t} + \frac{{\widehat
p}_{x0}}{m\omega}\sin{\omega t} \, , \nonumber\\
{\widehat p}_x (t) & = &  - m \omega {\widehat x}_0 \sin{\omega t} + {\widehat
p}_{x0} \cos{\omega t} \, ,\nonumber\\
{\widehat y} (t) & = & {\widehat y}_0 \cos{\omega t} + \frac{{\widehat
p}_{y0}}{m\omega}\sin{\omega t} \, , \nonumber\\
{\widehat p}_y (t) & = &  - m \omega {\widehat y}_0 \sin{\omega t} + {\widehat
p}_{y0} \cos{\omega t} \, .
\end{eqnarray}
So far, in the right-hand side of Eq. (\ref{QSol}) the operators ${\widehat
x}_0$, ${\widehat y}_0$, ${\widehat p}_{x0}$, and ${\widehat p}_{y0}$ are
``abstract", i.e., the {\it concrete} commutation relations satisfied by them
have {\it not}, at this stage, been specified. Moreover, the specification of
the algebraic relations satisfied by them gives rise precisely to distinct
quantum theories. Let ${\widehat O}(t=0)$ be any of the fundamental operators
${\widehat x}_0$, ${\widehat y}_0$, ${\widehat p}_{x0}$, or ${\widehat
p}_{y0}$. The corresponding quantum theories emerging from each of the
symplectic structures are the following:

0) The quantum theory emerging from the triple $(\Gamma=\mathbb{R}^4,
\omega=\omega_0, H= S_0)$ is defined by a representation of the
algebra
\begin{eqnarray}\label{QCR}
[ {\widehat x}_0 , {\widehat p}_{x0} ] = i \hbar\, , \quad [ {\widehat y}_0
,{\widehat p}_{y0} ] = i \hbar \, ,
\end{eqnarray}
associated with Eq. (\ref{CanPB}). Let the Hilbert space be the space of
square-integrable functions in $\mathbb{R}^2$, ${\cal F}={\cal L}^2
(\mathbb{R}^2, d\mu= dx dy)$, then
\begin{eqnarray}\label{SRep}
{\widehat x}_0 & = & x \, , \quad {\widehat p}_{x0} = \frac{\hbar}{i}
\frac{\partial}{\partial x} \, , \nonumber\\
{\widehat y}_0 & = & y \, , \quad {\widehat p}_{y 0} =
\frac{\hbar}{i}
\frac{\partial}{\partial y} \, ,
\end{eqnarray}
is a Schr\"odinger or coordinate representation of the fundamental operators
which satisfies (\ref{QCR}). In addition, the relationship between the
coordinate and momentum basis can be obtained from
\begin{eqnarray}
{\widehat x}_0 \mid x,y \rangle & = & x \mid x,y \rangle\, , \nonumber\\
{\widehat y}_0 \mid x,y \rangle & = & y \mid x,y \rangle \, ,
\nonumber\\
{\widehat p}_{x0} \mid p_x , p_y \rangle & = & p_x \mid p_x ,p_y \rangle \, ,
\nonumber\\
\quad {\widehat p}_{y0} \mid p_x , p_y \rangle & = & p_y \mid p_x , p_y
\rangle \, ,
\end{eqnarray}
and Eq. (\ref{SRep}) from which, after
normalization,
\begin{eqnarray}\label{FT0}
\langle x,y \mid p_x , p_y \rangle & = & \langle x \mid p_x \rangle \langle y
\mid p_y \rangle \nonumber\\
& = & \frac{1}{2\pi \hbar} e^{i(x p_x + y p_y)/\hbar} \, .
\end{eqnarray}
Moreover, the representation of the operators given in Eq. (\ref{SRep}) is
unitarily equivalent to
\begin{eqnarray}\label{FirstR}
{\widehat x} (t) & = & x \cos{\omega t} + \frac{\hbar}{m\omega i}\sin{\omega
t} \frac{\partial}{\partial x}\, , \nonumber\\
{\widehat p}_x (t) & = &  - m \omega x \sin{\omega t} + \frac{\hbar}{i}
\cos{\omega t} \frac{\partial}{\partial x} \, ,\nonumber\\
{\widehat y} (t) & = & y \cos{\omega t} + \frac{\hbar}{m\omega i}\sin{\omega
t}
\frac{\partial}{\partial y} \, , \nonumber\\
{\widehat p}_y (t) & = &  - m \omega y \sin{\omega t} + \frac{\hbar}{i}
\cos{\omega t} \frac{\partial}{\partial y} \, ,
\end{eqnarray}
obtained by using
\begin{eqnarray}\label{Unitary}
{\widehat O}(t) = e^{{i{\widehat S}_0 t}/\hbar} {\widehat O} (t=0)
e^{{-i{\widehat S}_0 t}/\hbar}\, ,
\end{eqnarray}
and Eqs. (\ref{QCR}) and (\ref{SRep}), i.e., equation (\ref{FirstR}) is the
concrete version of Eq. (\ref{QSol}) after the use of Eq. (\ref{SRep}).

1) Similarly, the quantum theory built from the triple $(\Gamma=\mathbb{R}^4,
\omega=\omega_1, H= S_1)$ is defined by a representation of the algebra
\begin{eqnarray}\label{QCR2}
[ {\widehat x}_0 , {\widehat p}_{y0} ] = i \hbar\, , \quad [ {\widehat y}_0 ,
{\widehat p}_{x0} ] = i \hbar \, ,
\end{eqnarray}
in agreement with Eq. (\ref{CanPBi}). Let the Hilbert space be ${\cal F}={\cal
L}^2 (\mathbb{R}^2, d\mu= dx dy)$ then
\begin{eqnarray}\label{1Rep}
{\widehat x}_0 & = & x \, , \quad {\widehat p}_{x0} = \frac{\hbar}{i}
\frac{\partial}{\partial y} \, , \nonumber\\
{\widehat y}_0 & = & y \, , \quad {\widehat p}_{y0} = \frac{\hbar}{i}
\frac{\partial}{\partial x} \, ,
\end{eqnarray}
is a Schr\"odinger representation of the fundamental operators which satisfies
(\ref{QCR2}) [cf. Eq. (\ref{SRep})]. Again, the relationship between the
coordinate and momentum basis can be obtained from
\begin{eqnarray}
{\widehat x}_0 \mid x,y \rangle & = & x \mid x,y \rangle\, , \nonumber\\
{\widehat y}_0 \mid x,y \rangle & = & y \mid x,y \rangle \, , \nonumber\\
{\widehat p}_{x0} \mid p_x , p_y \rangle & = & p_x \mid p_x ,p_y \rangle \, ,
\nonumber\\
\quad {\widehat p}_{y0} \mid p_x , p_y \rangle & = & p_y \mid p_x , p_y
\rangle \, ,
\end{eqnarray}
and Eq. (\ref{1Rep}) from which, after
normalization,
\begin{eqnarray}\label{FT1}
\langle x,y \mid p_x , p_y \rangle & = & \langle x \mid p_y \rangle \langle y
\mid p_x \rangle \, \nonumber\\
& = & \frac{1}{2\pi \hbar} e^{i(x p_y + y p_x)/\hbar}
\end{eqnarray}
[cf. Eq. (\ref{FT0})]. Note that we have, in this case, something that might
be called a ``crossed Fourier transform"  in the sense that a packet sharped
in the $x$ direction spreads out in the $p_y$ direction (same for $y$ and
$p_x$). Moreover, by the Stone-von Neumann theorem the operators given in Eq.
(\ref{1Rep}) are unitarily equivalent to
\begin{eqnarray}\label{SecondR}
{\widehat x} (t) & = & x \cos{\omega t} + \frac{\hbar}{m\omega i}\sin{\omega
t} \frac{\partial}{\partial y}
\, , \nonumber\\
{\widehat p}_x (t) & = &  - m \omega x \sin{\omega t} + \frac{\hbar}{i}
\cos{\omega t} \frac{\partial}{\partial y} \, ,\nonumber\\
{\widehat y} (t) & = & y \cos{\omega t} + \frac{\hbar}{m\omega i}\sin{\omega
t}
\frac{\partial}{\partial x} \, , \nonumber\\
{\widehat p}_y (t) & = &  - m \omega y \sin{\omega t} + \frac{\hbar}{i}
\cos{\omega t} \frac{\partial}{\partial x} \, ,
\end{eqnarray}
obtained by using
\begin{eqnarray}\label{Unitary2}
{\widehat O}(t) = e^{i{\widehat S}_1 t/\hbar} {\widehat O} (t=0)
e^{-i{\widehat S}_1 t/\hbar}\, ,
\end{eqnarray}
and Eqs. (\ref{QCR2}) and (\ref{1Rep}), i.e., equation (\ref{SecondR}) is the
concrete version of Eq. (\ref{QSol}) in the present quantum theory.

2) In the case of the triple $(\Gamma=\mathbb{R}^4 , \omega = \omega_2 , H=
S_2)$ the quantum theory is defined by a representation of the algebra
\begin{eqnarray}\label{QCR3}
[ {\widehat x}_0 , {\widehat p}_{x0} ] = - i \hbar\, , \quad [ {\widehat y}_0
, {\widehat p}_{y0} ] = i \hbar \, ,
\end{eqnarray}
in agreement with Eq. (\ref{CanPBii}). Let the Hilbert space be ${\cal
F}={\cal L}^2 (\mathbb{R}^2, d\mu= dx dy)$ then
\begin{eqnarray}\label{2Rep}
{\widehat x}_0 & = & x \, , \quad {\widehat p}_{x0} = - \frac{\hbar}{i}
\frac{\partial}{\partial x} \, , \nonumber\\
{\widehat y}_0 & = & y \, , \quad {\widehat p}_{y0} =
\frac{\hbar}{i}
\frac{\partial}{\partial y} \, ,
\end{eqnarray}
is a Schr\"odinger representation of the fundamental operators which satisfies
(\ref{QCR3}) [cf. Eq. (\ref{SRep})]. Once again, the relationship between the
coordinate and momentum basis can be obtained from
\begin{eqnarray}
{\widehat x}_0 \mid x,y \rangle & = & x \mid x,y \rangle\, , \nonumber\\
{\widehat y}_0 \mid x,y \rangle & = & y \mid x,y \rangle \, , \nonumber\\
{\widehat p}_{x0} \mid p_x , p_y \rangle & = & p_x \mid p_x ,p_y \rangle \, ,
\nonumber\\
\quad {\widehat p}_{y0} \mid p_x , p_y \rangle & = & p_y \mid p_x , p_y
\rangle \, ,
\end{eqnarray}
and Eq. (\ref{2Rep}) from which, after normalization,
\begin{eqnarray}\label{FT2}
\langle x,y \mid p_x , p_y \rangle & = & \langle x \mid p_x \rangle \langle y
\mid p_y \rangle \nonumber\\
& = & \frac{1}{2\pi \hbar} e^{i(- x p_x + y p_y)/\hbar}
\end{eqnarray}
[cf. Eq. (\ref{FT0})]. As expected, on account of the Stone-von Neumann
theorem, the operators given in Eq. (\ref{2Rep}) are unitarily equivalent to
\begin{eqnarray}\label{ThirdR}
{\widehat x} (t) & = & x \cos{\omega t} - \frac{\hbar}{m\omega i}\sin{\omega
t} \frac{\partial}{\partial x} \, , \nonumber\\
{\widehat p}_x (t) & = &  - m \omega x \sin{\omega t} - \frac{\hbar}{i}
\cos{\omega t} \frac{\partial}{\partial x} \, ,\nonumber\\
{\widehat y} (t) & = & y \cos{\omega t} + \frac{\hbar}{m\omega i}\sin{\omega
t}\frac{\partial}{\partial y} \, , \nonumber\\
{\widehat p}_y (t) & = &  - m \omega y \sin{\omega t} + \frac{\hbar}{i}
\cos{\omega t} \frac{\partial}{\partial y} \, ,
\end{eqnarray}
with the unitary transformation given by
\begin{eqnarray}\label{Unitary3}
{\widehat O}(t) = e^{i{\widehat S}_2 t/\hbar} {\widehat O} (t=0)
e^{-i{\widehat S}_2 t/\hbar}\, ,
\end{eqnarray}
and taking into account Eqs. (\ref{QCR3}) and (\ref{2Rep}), i.e., equation
(\ref{ThirdR}) is the concrete version of Eq. (\ref{QSol}) in the present
case.

3) finally, the quantum theory associated with the triple
$(\Gamma=\mathbb{R}^4 , \omega = \omega_3 , H= S_3)$ is built from a
representation of the algebra
\begin{eqnarray}\label{QCR4}
[ {\widehat x}_0 , {\widehat y}_{0} ] = - \frac{i \hbar}{m\omega} \, , \quad [
{\widehat p}_{x0} , {\widehat p}_{y0} ] = - i \hbar m \omega \, ,
\end{eqnarray}
in agreement with Eq. (\ref{CanPBiii}). However, this case is a little bit
different from the cases 0), 1), and 2). There, independently of the case, the
operators ${\widehat p}_{x0}$ and ${\widehat p}_{y0}$ commute. This is also
the case of the operators ${\widehat x}_0$ and ${\widehat y}_0$. This fact was
used to build a ``coordinate" and ``momentum" basis and their interconnection
was displayed in all cases. But now, ${\widehat p}_{x0}$ and ${\widehat
p}_{y0}$ do not commute anymore (same for the operators ${\widehat x}_0$ and
${\widehat y}_0$). Thus, in the present case, it is not possible to build a
common basis for these operators as before. Nevertheless, it makes sense to
talk about a Schr\"odinger or ``coordinate representation" for the operators
involved. By this, we mean
\begin{eqnarray}\label{4Rep}
{\widehat x}_0 & = & x \, , \quad \quad \quad \quad {\widehat p}_{x0} =
m\omega y \, , \nonumber\\
{\widehat y}_0 & = & \frac{i \hbar}{m\omega} \frac{\partial}{\partial x} \, ,
\quad \,\, {\widehat p}_{y0} = i \hbar \frac{\partial}{\partial y} \, ,
\end{eqnarray}
which satisfies Eq. (\ref{QCR4}). This representation for the operators is, by
means of the Stone-von Neumann theorem, unitarily equivalent to
\begin{eqnarray}\label{FourthR}
{\widehat x} (t) & = & x \cos{\omega t} + y \sin{\omega t} \, , \nonumber\\
{\widehat p}_x (t) & = & - m \omega x \sin{\omega t} + m \omega y \cos{\omega
t}\, , \nonumber\\
{\widehat y} (t) & = & \frac{i \hbar}{m\omega} \cos{\omega t}
\frac{\partial}{\partial x} + \frac{i \hbar}{m\omega } \sin{\omega t}
\frac{\partial}{\partial y} \, , \nonumber\\
{\widehat p}_y (t) & = & - i \hbar \sin{\omega t} \frac{\partial}{\partial x}
+ i \hbar \cos{\omega t} \frac{\partial}{\partial y} \, ,
\end{eqnarray}
via
\begin{eqnarray}\label{Unitary4}
{\widehat O}(t) = e^{i{\widehat S}_3 t/\hbar} {\widehat O} (t=0)
e^{-i{\widehat S}_3 t/\hbar} \, ,
\end{eqnarray}
and Eqs. (\ref{QCR4}) and (\ref{4Rep}), i.e., equation (\ref{FourthR}) is the
concrete version of Eq. (\ref{QSol}) in this case.

{\it Inequivalence of the quantum theories}. So far, four mathematically
consistent quantum theories have been obtained by using Dirac's quantization
rule, which is a cornerstone of quantum mechanics. In each of these theories,
evolution in $t$ is a unitary transformation. Now, according to Heisenberg's
picture of quantum mechanics if the system is left (prepared) on by means of a
certain experimental arrangement in the state $\mid\Psi \rangle$ (which might
be even a wave packet) at $t=0$ then
\begin{eqnarray}\label{meanvalues}
\langle \Psi \mid {\widehat O}(t) \mid \Psi \rangle
\end{eqnarray}
yields the expected (central) value in the distribution of the corresponding
physical quantity associated with the observable ${\widehat O}(t)$ if that
quantity were to be measured at time $t$. At first sight it might appear that
the numerical value of the expectation value (\ref{meanvalues}) for certain
(and fixed) observable ${\widehat O}(t)$ is the same in all the four quantum
theories under consideration, after all Eq. (\ref{QSol}) which is required to
compute (\ref{meanvalues}) has, apparently, the same functional form for all
of these theories. However, this is {\it not} so for the simple reason that in
each of the quantum theories described above the fundamental operators
${\widehat x}_0$, ${\widehat y}_0$, ${\widehat p}_{x0}$, and ${\widehat
p}_{y0}$ {\it act very differently} on the state $\mid \Psi \rangle$ in which
the system was prepared on because such operators have quite distinct
representations on account of the specific algebraic relations they must
satisfy in each theory. For instance
\begin{eqnarray}
&  & x \cos{\omega t} - \frac{i\hbar}{m\omega} \sin{\omega t}
\frac{\partial}{\partial x}\, , \nonumber\\
&  & x \cos{\omega t} - \frac{i \hbar}{m\omega} \sin{\omega t}
\frac{\partial}{\partial y}\, , \nonumber\\
&  & x \cos{\omega t} + \frac{i\hbar}{m\omega} \sin{\omega t}
\frac{\partial}{\partial x}\, , \nonumber\\
&  & x \cos{\omega t} + y
\sin{\omega t} \, ,
\end{eqnarray}
are the corresponding operators associated to the observable ${\widehat x}(t)$
in the quantum theories 0), 1), 2), and 3); respectively. Therefore, the
various quantum theories are {\it inequivalent} in the sense that the
expectation value (\ref{meanvalues}) of the fundamental operators ${\widehat
O}(t)$ computed by using one quantum theory is not the same expectation value
than the one obtained with any other of the quantum theories analyzed above
when the system is prepared in the state $\mid \Psi \rangle$ (same for all
theories). Note that this inequivalence between the various quantum theories
does {\it not} arise from an ambiguity in the order of the operators as
usually happens when there are several quantum theories associated to a single
classical theory. The origin of the inequivalence comes from: 1) the various
quantum theories emerging from the implementation of Dirac's quantization rule
to the several symplectic structures chosen to make the classical description
and 2) keeping the interpretation that the state in which the system is
prepared on by the experimental arrangement has the same functional form in
all the quantum theories.

From the previous discussion it is clear that, at this stage, theoretical
consistency in the construction of the quantum theories does not provide a
unique way of relating theoretical predictions with experimental outcomes.
Either:

a) nature prefers just one of the various quantum theories in the sense that
only one of these quantum theories matches the experimental data. Even if this
were the case, there would still be something missing in the theoretical
formalism whose knowledge might allow us to pick up a particular quantum
theory and discard the remaining ones solely on theoretical grounds, i.e., we
would need to specify that hypothetic rule that would allow us to single out
the ``right" quantum theory and also to uncover the fundamental cause of this,
or

b) all the quantum theories are mathematically and physically viable. From
this perspective, one would be assuming that there should exist a (yet
unknown) {\it covariant quantization scheme} without the need of restricting
ourselves to the use of a particular symplectic structure as starting point to
build the quantum theory. Nevertheless, due to the fact that the expected
values computed in each theory are numerically distinct, this would mean that
there should exist a (yet unknown) criterion whose knowledge and its
implementation would lead to the same theoretical predictions (which will
match the experimental data) no matter which symplectic structure were chosen
from the very beginning.

\section{Heisenberg's uncertainty principle and measuring process}\label{Hup}
The consequences of having various quantum theories built from the
implementation of Dirac's quantization rule to the various symplectic
structures for the fundamental variables $x$, $y$, $p_x$, and $p_y$ are much
more stronger when the several uncertainty relations coming from such
quantization schemes are analyzed. To appreciate this, it is convenient to
remind the reader that the {\it physical meaning} of the observables
${\widehat x}(t)$, ${\widehat y}(t)$, ${\widehat p}_{x0}(t)$, and ${\widehat
p}_{y0}$ {\it is the same} in spite of the specific representation the
operators acquire in each one of the theories under study.

The non-trivial products of quantum uncertainties in the measurement of
${\widehat x(t)}$, ${\widehat y}(t)$, ${\widehat p}_x (t)$, and ${\widehat
p}_y (t)$ are, in each theory, given by:

0)
\begin{eqnarray}
\Delta x \Delta p_x \geq \frac{\hbar}{2} \, , \quad \Delta y \Delta p_y \geq
\frac{\hbar}{2} \, . \label{Uncer0}
\end{eqnarray}

1)
\begin{eqnarray}\label{Uncer1}
\Delta x \Delta p_y \geq \frac{\hbar}{2} \, , \quad \Delta y \Delta p_x \geq
\frac{\hbar}{2} \, .
\end{eqnarray}

2)
\begin{eqnarray}\label{Uncer2}
\Delta x \Delta p_x \geq \frac{\hbar}{2} \, , \quad \Delta y \Delta p_y \geq
\frac{\hbar}{2} \, .
\end{eqnarray}

3)

\begin{eqnarray}
\Delta x \Delta y \geq \frac{\hbar}{2m\omega}\, , \quad \Delta p_x \Delta p_y
\geq \frac{\hbar m \omega}{2} \, . \label{Uncer3}
\end{eqnarray}
Once again, from a) and b) of Sect. \ref{quantum} either nature prefers a
single quantum theory or the sets of product of uncertainties given in Eqs.
(\ref{Uncer0})-(\ref{Uncer3}) are just a reflection of the fact the standard
uncertainty relation is {\it not} expressed in a covariant way.

Moreover, from Eq. (\ref{QSol}) one has, just to list an example
\begin{eqnarray}
\left [ {\widehat x}(t) , {\widehat x}(t') \right ] & = & \frac{1}{m\omega}
\sin{\omega(t' -t)} \left [ {\widehat x}_0 , {\widehat p}_{x0} \right ] \, ,
\end{eqnarray}
which means, according to the standard interpretation of quantum mechanics,
that the variable ${\widehat x(t)}$ can be monitored without affecting its
evolution in the framework of theories 1) and 3) but not in the quantum
theories 0) and 2) [see page 380 of Ref. \cite{Peres}].

\section{Discussion}\label{Remarks}
In the symplectic viewpoint of dynamics it is possible to make the description
of a dynamical system without having the necessity of restricting ourselves to
the case where $(q^i,p_i)$ are canonical pairs. The various quantum theories
built from the application of Dirac's quantization rule to these alternative
symplectic structures yields inequivalent quantum theories in the sense that
the expectation values for observables representing the same physical quantity
are different. However, we think that it should be possible to build a
quantization scheme which matches experimental outcomes no matter if the
$(q^i,p_i)$ are or not canonical. After all, nature should not care which type
of symplectic structure one uses to describe it. Experimental results should
be independent of each particular choice of symplectic structure. Thus, our
phylosophical position is closer to the point b) of Sect. \ref{quantum}.
Finally, the consequences of choosing different pairs $(\omega,H)$ formed by a
symplectic structure $\omega$ and a Hamiltonian $H$ in field theory (where
expansion on harmonic oscillators is frequently done) as well as on classical
and quantum statistical mechanics are not explored, but deserve to be done.
Also, the possibility of choosing alternative symplectic theories in realistic
theories such as general relativity or string theories and the consequences of
this fact on their quantum theories should be investigated.

\section*{Acknowledgements}
Warm thanks to Abdel P\'erez-Lorenzana for very fruitful discussions on the
topics of this paper.


\end{document}